\documentclass[a4paper,11pt]{article}
\usepackage{pos}

\title{A Portal Dedicated to Higgs Bosons for Experts and the General Public}
\author*[a,b]{Andr\'e Sopczak}
\affiliation[a]{Czech Technical University in Prague, Institute of Experimental and Applied Physics,\\
Husova 240/5, CZ-110\,00, Prague 1, Czechia}
\affiliation[b]{on behalf of the International Particle Physics Outreach Group}

\emailAdd{andre.sopczak@cern.ch}

\abstract{As an educational aid and 
source for expert information,
 a web portal dedicated to Higgs boson research is presented. 
A database is created with more than 1000 relevant articles using CERN Document Server API and web scraping methods. 
The database is automatically updated when new results on the Higgs boson become available. 
Using artificial intelligence and natural language processing, 
the articles are categorized according to properties of the Higgs boson and other criteria. 
The process of designing and implementing the Higgs Boson Portal (HBP) is described. 
The components of the HBP are deployed to CERN Web Services using the OpenShift cloud platform. 
The HBP is accessible within the Czech Particle Physics Project (CPPP) at 
http://cern.ch/cppp and directly at http://cern.ch/higgs.}

\FullConference{%
    41st International Conference on High Energy physics - ICHEP2022\\
  6-13 July, 2022\\
  Bologna, Italy
}

\begin{document}
\maketitle

\section{Introduction}

Over the past decades, the search for the Higgs boson of the Standard Model and Higgs bosons in extended models have been at the forefront of research in particle physics. Ten years ago the discovery of a Higgs boson with Standard Model properties was established.
The Higgs boson research has led to more than 1000 experimental publications on the Higgs boson search and the measurement of its properties. Many aspects of the Higgs boson research have been addressed, including the search for Higgs bosons beyond the Standard Model.
For the benefit of experts and the general public interested in this exciting field of research, a Higgs Boson Portal (HBP) is implemented as
part of the Czech Particle Physics Project together with
other modules~\cite{Modules_these_proc}.

The HBP uses 
Artificial Intelligence (AI) and
Natural Language Processing (NLP) for categorization of the Higgs boson publications in a user-accessible list with the following categorizations:
\begin{itemize}
\item Publication stage (preprint, journal accepted, published)
\item Year of public release
\item Experiment
\item Luminosity 
\item Higgs boson decay mode
\item Higgs boson production mode
\item Higgs boson model (Standard Model or Beyond Standard Model)
\end{itemize}
The detailed categorization 
of the production and decay modes 
follows the review~\cite{Sopczak:2020vrs}.

The database is automatically updated when new results on the Higgs boson become available, and
the components of the HBP are deployed to CERN Web Services.
%using the OpenShift cloud platform. 
%
The HBP also includes a visualisation of some developments of limits and precision.

Following a feasibility study~\cite{Kupka:2722144},
implementation~\cite{Zacik:2774895} and 
presentation~\cite{ippog2021},
the HBP is accessible on http://cern.ch/higgs.

%\clearpage
\section{Sources of Publications}

The publication information is extracted using web scraping of the Fermilab web sites~\cite{CDF,D0} and 
with the CERN Document Server API~\cite{CDS}.
Publications of the following collaborations 
are included:
\begin{itemize}
    \item 1989 – 2000: CERN – Large Electron-Positron Collider (LEP)
    \begin{itemize}
    \item ALEPH, DELPHI, L3, OPAL
    \end{itemize}
    \item 1987 – 2011 Fermilab – Tevatron Collider
    \begin{itemize}
    \item CDF, D0
    \end{itemize}
    \item 2010 – present: CERN – Large Hadron Collider (LHC)
    \begin{itemize}
    \item ATLAS, CMS
    \end{itemize}
\end{itemize}

\section{Natural Language Processing}

An expert in the field of Higgs boson research recognizes immediately the category of an article.
However, this identification is not a 
trivial task for an automated computer categorization.
Therefore, a complex solution is chosen 
applying Artificial Intelligence (AI) and 
Natural Language Processing (NLP).
Details and further references are given in Ref.~\cite{Kupka:2722144}.
A probability for a publication is assigned for all the predefined classes. The class with the highest probability is chosen for the categorization.

A publication text is expressed as a vector of words. %This is commonly called the bag of words approach [23]. 
Before classification, it is beneficial to 
pre-process the text by removing stopwords, which are the most frequently used words in the given language. English examples are “the”, “and” and “from”. 
These words do not play any role in classifying an article. 
The remaining words undergo stemming, 
which involves reducing the words to their root form, e.g. “decaying” to “decay”. 
Although the context of individual words 
is lost during the pre-processing, this method can still be used to classify the text as a whole.
It is computationally ineffective
to apply NLP to the full text of the article. Therefore, for the HBP, only the title and the 
abstract of an article are selected for processing.

The HBP utilises a naive Bayes classifier to distinguish between articles studying 
the Standard Model and articles searching beyond the Standard Model. 
These two directions of research generally
use slightly different wording, which helps the classification process to be accurate.
For example, the word “search” appears more often 
in publications beyond the Standard Model.

%The process of automatically recognising and 
%extracting meaningful words, 
%phrases or numeric values is called 
%Named Entity Recognition (NER).

In order to extract more concrete information 
from the articles, e.g. integrated luminosity,
centre-of-mass energy, decay products or production modes, a different method is applied.
These entities are dependent on the context in
which they are used.
For example~\cite{Aaboud:2636382}, in the title: 
“Measurements of gluon fusion and vector-boson-fusion production of the Higgs boson in 
$H\rightarrow WW^* \rightarrow e\nu\mu\nu$
decays using pp collisions at $\sqrt s = 13$\,TeV 
with the ATLAS detector”, two
production modes are mentioned – gluon fusion and vector boson fusion. 
The decay mode is expressed using the notation 
“$H\rightarrow WW^*$”. 
The centre-of-mass
energy ($\sqrt s$) is given as the number “13” 
combined with the “TeV” unit.
The process of automatically recognising and 
extracting meaningful words,
phrases or numeric values is called Named Entity Recognition (NER).
NER has to recognize syntactic structures of the text, and therefore is language specific. 
Most scientific articles are written in English, therefore the HBP uses a NER model pre-trained on the English language. 
The training can be extended for specific 
needs of the domain.
After extracting entities from the text, they have to be parsed or categorized.
When numeric values are expected, the number and unit are parsed algorithmically using 
a set of predetermined rules. 
In case of the decay mode and the production mode, 
the category is decided by identifying keywords
and special characters in the extracted named entity. %Identifying keywords and special characters 
%only in the named entity yields better results that
%searching in the full text, where the context of 
%the keywords are ignored.
%
The performance of the NER categorization
is given in Table~\ref{tab:performance}.

\begin{table}[hbp]
    \begin{center}
    \caption{NER categorization performance.
    True Positives (TP), False Positives (FP),
    False Negatives (FN).}
        \label{tab:performance}
        %\scalebox{0.8}[0.9]{
            \begin{tabular}{c|cccccc}
                & TP & FP & FN & Precision (\%) & 
                    Recall (\%) & $F_1$-score (\%) \\ \hline
                Luminosity  &51&2&7&96.2&87.9&91.9 \\ 
                Centre-of-mass energy &51&0&9&100&85.0&91.9 \\
                Production mode &52&8&9&86.7&85.3&86.0\\
                Decay mode &71&17&19&80.7&78.9&79.8
        \end{tabular}
        %}
    \end{center}
    \vspace{-7mm}
\end{table}

\clearpage

\section{User interface}
\begin{figure}[!hbtp]
    \vspace*{-2mm}
    \centering
    \includegraphics[width=\textwidth]{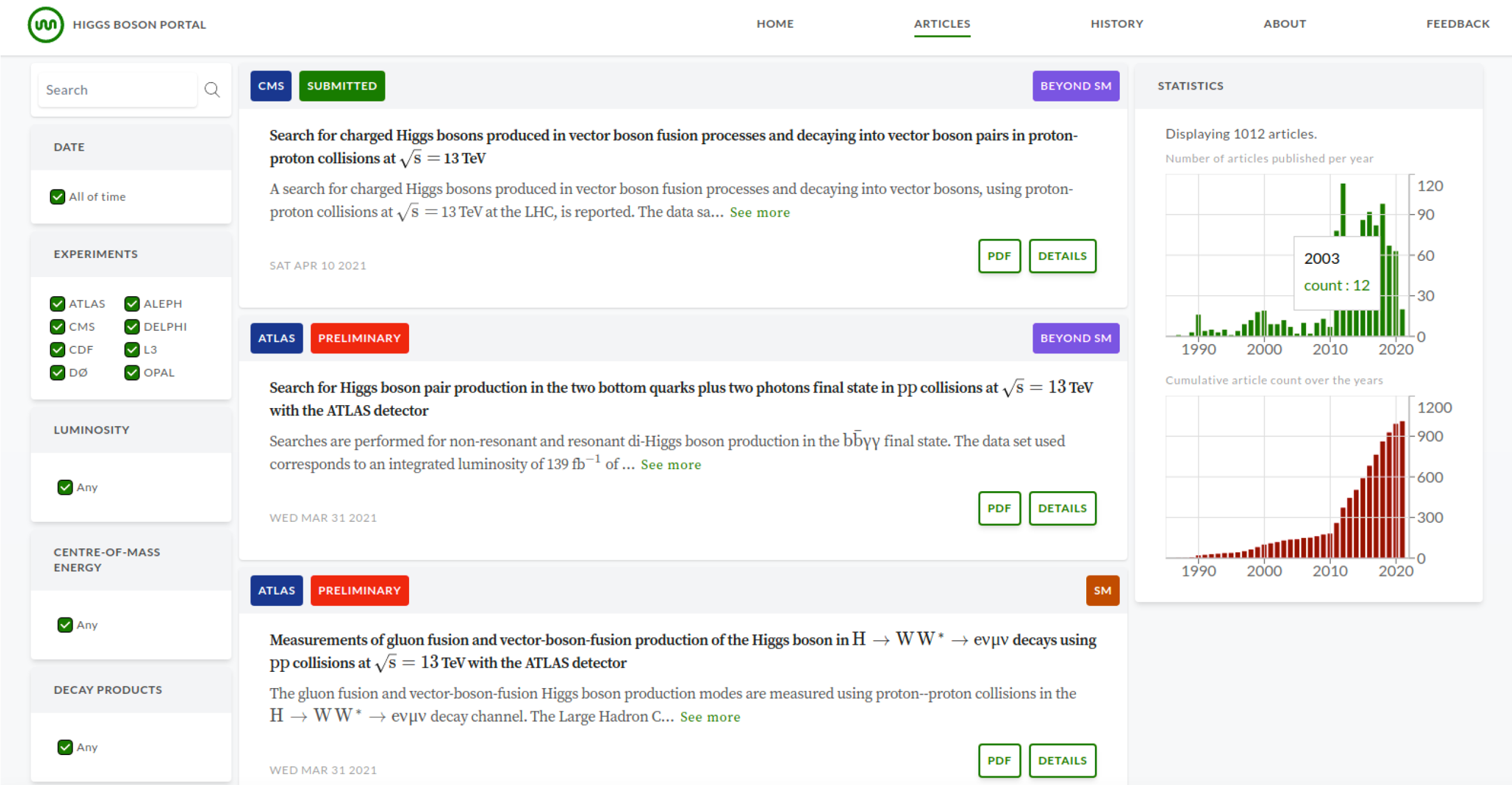}
    \vspace*{-7mm}
    \caption{User interface Higgs Boson Portal (HBP).}
    \label{fig:interface}
    \vspace*{-4mm}
\end{figure}

\vspace*{-2mm}
The user interface is shown in Fig.~\ref{fig:interface}.

\section{Developments of Limits and Measurements}

\begin{figure}[!hbtp]
    \vspace*{-2mm}
    \centering
    \includegraphics[width=0.32\textwidth]{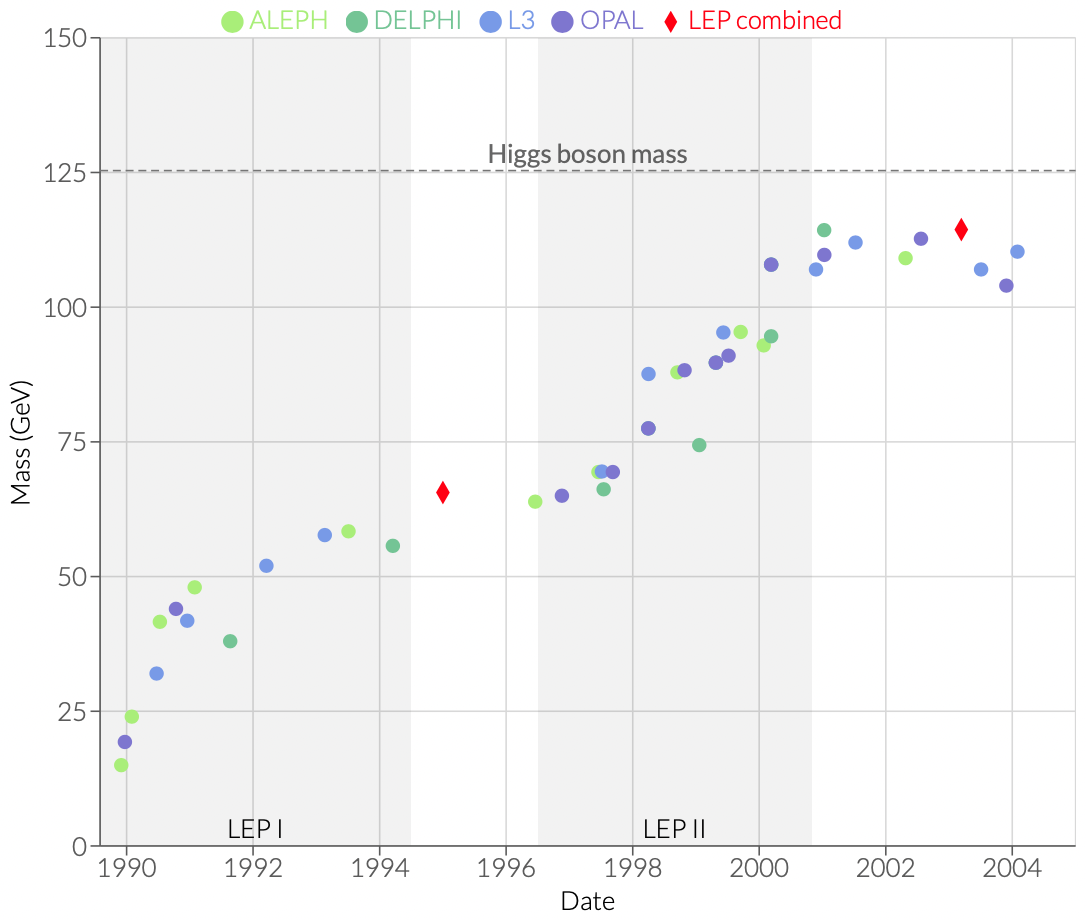}
    \includegraphics[width=0.32\textwidth]{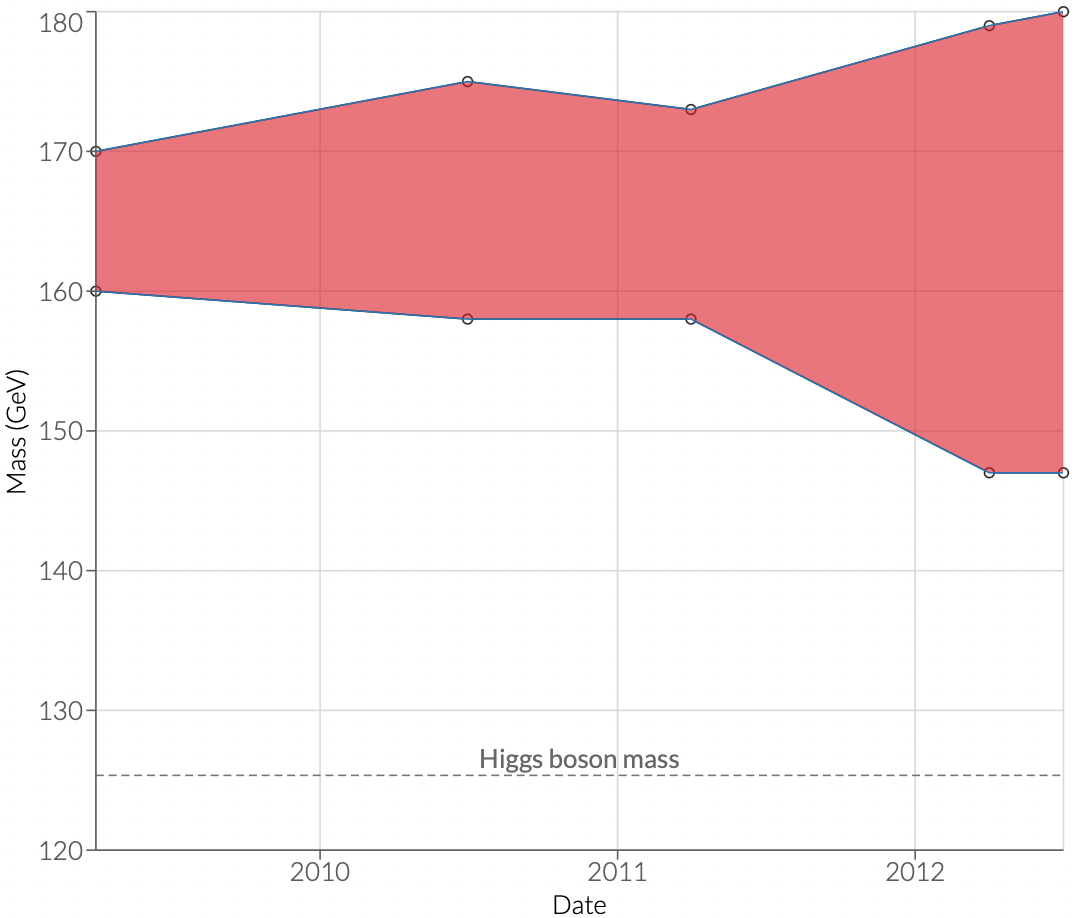}
 \includegraphics[width=0.32\textwidth]{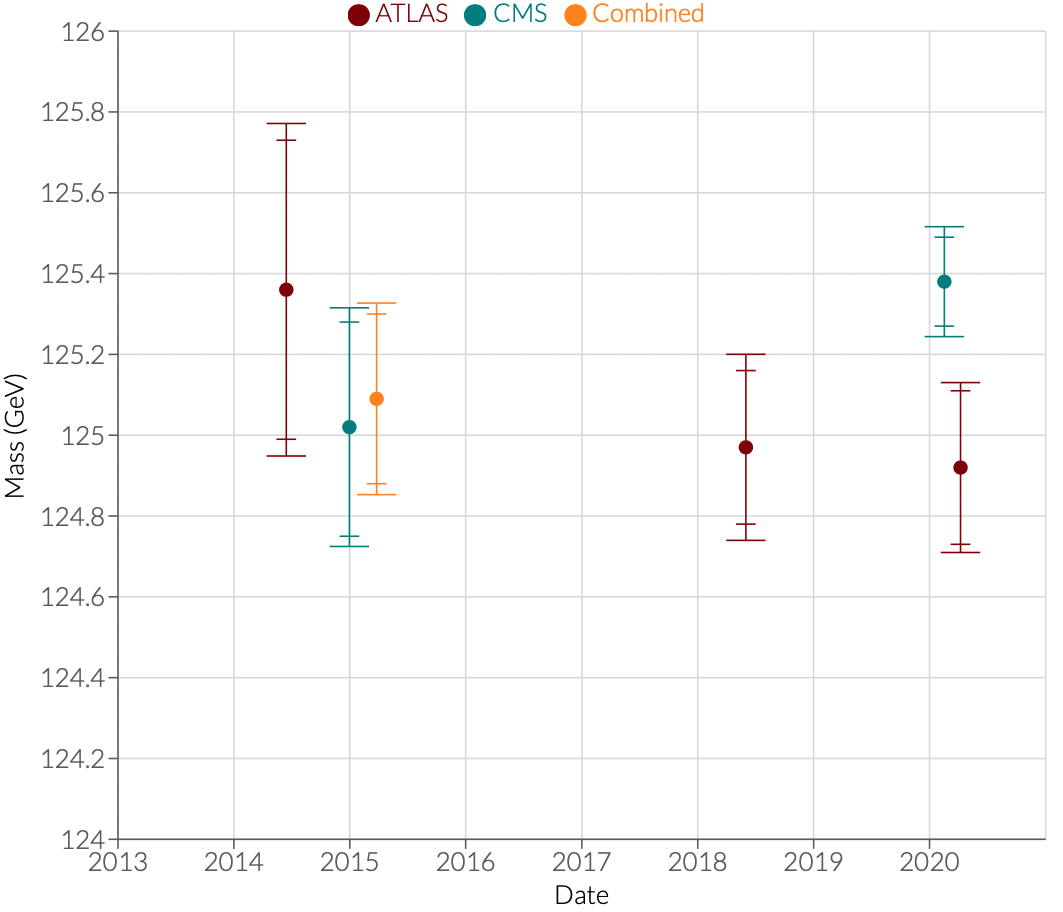}   
    \vspace*{-2mm}
    \caption{Developments of limits and 
measurements in the Higgs boson research.}
    \label{fig:history}
    \vspace*{-5mm}
\end{figure}

The "History" link in the user interface gives 
examples of the developments of limits and 
measurements in the Higgs boson research (Fig.~\ref{fig:history}). Clicking on a data point leads to the corresponding publication, 
and the development of the upper Higgs boson 
mass limit is taken from Ref.~\cite{Sopczak_2012}.

\section{Administration and Maintenance}

The user interface allows administration login. 
The administrator can adjust the NLP categorization.
Categorized publications are stored in a Mongo database.
Updates are performed daily with Python cron jobs.

\section{Conclusion}
A Higgs Boson Portal has been established with more than 1000 categorized publications and daily updates. It serves as an educational aid and data base for experts.

\clearpage
\section{Acknowledgements}
The author would like to 
thank the 
International Particle Physics Outreach Group
(IPPOG) for fruitful discussions, and the students
Martin Kupka, Antoine Vauterin and Peter Zacik
for their dedication.
The grant support from the "Fondu celoškolských aktivit pro rok 2022" Czech Technical University in Prague is gratefully acknowledged.
The project is also supported by the Ministry of
Education, Youth and Sports of the Czech
Republic under the project number
LTT 17018.

\bibliographystyle{JHEP}
\bibliography{biblio}

%\begin{thebibliography}{99}
%\bibitem{...}
%\end{thebibliography}

\end{document}